# Leveraging Conformational Diversity for Enhanced Structure-Based Virtual Screening: Insights from Molecular Dynamics Simulations of HIV-1 Protease–Ligand Complexes

Pei-Kun Yang

E-mail: peikun@isu.edu.tw

**Abstract**

Structure-based virtual screening aims to identify high-affinity ligands by estimating binding free energies between proteins and small molecules. However, the conformational flexibility of both proteins and ligands challenges conventional rigid docking methods that assume a fixed receptor structure. In this study, we examined the impact of conformational diversity on binding energy calculations using 79 HIV-1 protease–ligand complexes. Molecular dynamics simulations were employed to generate structural ensembles for both proteins and ligands in aqueous environments. RMSD-based clustering was applied to reduce redundancy while preserving structural diversity. Binding energies were computed using van der Waals and electrostatic interactions. The results demonstrated that native protein–ligand pairs consistently yielded favorable binding energies, whereas non-native pairings often failed to reproduce binding. Furthermore, clustering thresholds influenced the balance between computational cost and interaction accuracy. These findings underscore the importance of incorporating multiple protein and ligand conformations in SBVS protocols to improve prediction reliability and support more effective drug discovery strategies.

# Introduction

Structure-based virtual screening (SBVS) requires the computation of binding free energies between proteins and ligands to identify potential compounds with high affinity for a given protein target. Even considering only low-molecular-weight compounds with drug-like potential, the chemical space can be as vast as $10^{60}$ molecules [1]. Current databases, such as ZINC-22, list over $10^{10}$ purchasable ligands [2], and the Enamine REAL Space encompasses approximately $6.5\times10^{10}$ synthetically accessible compounds [3], all of which may exhibit potential biological activity. Given the enormous number of molecules, it is impractical to experimentally evaluate the inhibitory effects of each compound on a target protein. Thus, virtual screening methods are employed to narrow down the candidate pool, thereby enhancing the efficiency of drug discovery. [4-6].

SBVS typically estimates binding free energies or employs scoring functions based on the structures of protein-ligand complexes [7-9]. However, the available structural data often represent the protein in an aqueous solution or a complex with a specific ligand rather than providing accurate complex structures for every potential ligand in a database [10]. Binding induces conformational adjustments in both the protein and the ligand to minimize the overall free energy, resulting in distinct conformational states when a protein binds different ligands and vice versa. Therefore, SBVS requires the generation of multiple possible conformations for both proteins and ligands, with the expectation that these conformations collectively encompass the binding modes observed in actual complexes. Subsequent binding free energy calculations are then performed for all combinations of these pre-generated conformations [11, 12].

To obtain the relevant conformations, a known protein structure is often used as the starting point to generate alternative conformations. Common approaches include side-chain rotations, which create diverse binding pocket conformations by altering dihedral angles; however, this method faces challenges in excluding sterically unfavorable conformations and does not typically account for backbone movement. Alternatively, molecular dynamics (MD) simulations can explore the conformational landscape of proteins [13-16]. Although MD simulations avoid generating unrealistic conformations, the distribution of states follows a Boltzmann distribution, necessitating longer simulation times to sample higher-energy conformations and to overcome energy barriers.

For ligands, a wide range of three-dimensional conformations is available from databases such as the RCSB PDB Ligand Expo, ZINC Database, PubChem, DrugBank, ChEMBL, and the Cambridge Structural Database [2, 10, 17-19]. These experimentally determined conformations can serve as initial structures, while tools like RDKit and OpenBabel can generate possible ligand conformations [20, 21]. Ligand conformations can be generated through systematic rotations around rotatable bonds or via molecular dynamics (MD) simulations. However, these approaches require careful filtering to exclude sterically unfavorable conformers and strategies to overcome high energy barriers.

This study employed seventy-nine HIV-1 protease-ligand complex structures to explore the influence of protein and ligand conformational variations on calculating binding energies. This provided insights into the optimal generation and selection of conformational ensembles for structure-based virtual screening.

**Method**

**Molecular Dynamics Simulations.** Seventy-nine HIV-1 protease–ligand complex structures were retrieved from the RCSB Protein Data Bank (PDB), with identical protein sequences but distinct bound ligands. The apo form of HIV-1 protease (PDB ID: 3IXO), which lacks any bound ligand, was selected based on sequence identity.

Input files for molecular dynamics simulations were generated using the CHARMM-GUI interface, employing the all36m_prot and all36_cgenff force fields [22-27]. All systems were solvated using the CHARMM-modified TIP3P water model [28, 29], and counterions ($Na^+$/$Cl^-$) were added to neutralize the net charge. NBFIX parameters were applied to improve specific ion–ligand interactions, in accordance with CHARMM force field recommendations. All simulations were performed using OpenMM [30]. Protonation states of titratable residues and ligands were assigned based on standard protonation patterns at pH 7.

Each of the 79 HIV-1 protease–ligand complexes was subjected to 5,000 steps of energy minimization, and the minimized structures were used for subsequent analysis; no equilibration or production simulations were performed for these complexes. The apo HIV-1 protease structure (PDB ID: 3IXO) was minimized for 5,000 steps, followed by equilibration at 303.15 K for 125 picoseconds using a 1-femtosecond timestep. A 100-nanosecond production simulation was then performed at 303.15 K using a 2-femtosecond timestep. For the ligands extracted from the complexes, each was placed in a water box with counterions, then underwent 5,000 steps of energy minimization, equilibration at 303.15 K for 125 picoseconds (1-fs timestep), and a 2-nanosecond production simulation (2-fs timestep). Trajectory frames were saved every 1 ps for the apo protein and every 0.2 ps for the ligands, resulting in $10^5$ and $10^4$ frames, respectively.

During the equilibration phase, van der Waals interactions were computed using a potential-switching function between 10 Å and 12 Å, and truncated beyond 12 Å. Electrostatic interactions were calculated using the particle-mesh Ewald method with a 12 Å real-space cutoff. A dielectric constant of 1.0 was applied. The SHAKE algorithm was used to constrain bonds involving hydrogen atoms, allowing a 2-fs integration timestep. The system was maintained at 303.15 K and 1 atm pressure using standard temperature and pressure coupling methods.

**Identification of Binding Pocket Residues.** Amino acids frequently interacting with the ligand were identified to define the binding pocket residues. Residues within 3 Å of the ligand were extracted for each of the 79 HIV-1 protease-ligand complexes. The frequency of occurrence for each residue across all complexes was statistically analyzed, and the most frequently observed residues were selected as binding pocket residues.

**Superimposition.** The equilibrated structure of PDB ID: 3IXO was used as the reference for structural alignment. A two-step superimposition process was applied to the protein and ligand components to maintain consistency across MD trajectories. The trajectory frames of PDB ID: 3IXO and the 79 HIV-1 protease-ligand complex structures were aligned to the reference structure by superimposing the Cα atoms of the binding pocket residues [31]. This ensured that the binding pocket remained structurally consistent across simulations. The trajectory frames of the 79 ligands were individually superimposed onto the ligand positions of their respective complex structures, aligning all non-hydrogen atoms to retain spatial positioning within the binding pocket.

**Clustering.** RMSD-based clustering was performed separately for protein and ligand conformations to reduce computational complexity while preserving structural diversity. For proteins, clustering was applied to the trajectory frames of the apo structure (PDB ID: 3IXO). Prior to clustering, all frames were aligned to the reference structure using the Cα atoms of the binding pocket residues. RMSD values were then computed using all heavy atoms of the binding pocket. Hierarchical clustering was performed using a range of RMSD thresholds ($RMSD_{thr}$) to group similar protein conformations and eliminate redundancy. For ligands, each of the 79 ligand trajectories was clustered independently. Before RMSD calculation, each frame was aligned to the corresponding bound pose using all non-hydrogen atoms. Hierarchical clustering based on all-atom RMSD was then applied to remove redundant ligand poses and retain distinct conformers. RMSD thresholds were systematically evaluated to assess their effect on conformational selection.

**Calculation of Electrostatic and vdW Interactions.** Electrostatic ($E_{ele}$) and van der Waals (vdW, $E_{vdw}$) interaction energies between the protein and ligand were computed using force field parameters from top_all36_prot and top_all36_cgenff, with the total binding energy defined as $E_{net} = E_{ele} + E_{vdw}$. $E_{ele}$ was calculated based on atomic partial charges, while $E_{vdw}$ was determined using vdW parameters from the force field. Only residues within the defined binding pocket were considered to ensure a precise evaluation of binding interactions, allowing for a focused and accurate assessment of protein-ligand interactions within the active site.

**Results**

**Binding Pocket Residues.** To simplify calculations, residues near the ligand were selected to define the binding pocket, including those from chains A and B. The identified residues were ARG 8, LEU 23, ASP 25, GLY 27, ALA 28, ASP 29, ASP 30, VAL 32, ILE 47, GLY 48, GLY 49, ILE 50, PRO 81, VAL 82, and ILE 84, totaling 30 residues.

These residues defined the binding pocket for subsequent protein-ligand binding energy calculations. $E_{ele}$ and $E_{vdw}$ were computed exclusively between these 30 binding pocket residues and the ligand, ensuring a focused analysis of key binding interactions.

**Protein Binding Pocket Specificity to Ligands.** To assess the necessity of multiple protein conformations in docking simulations, we retrieved 79 HIV-1 protease–ligand complexes from the PDB, where the protein sequences were identical, but the bound ligands varied. MD simulations were conducted to obtain energy-minimized protein conformations for each complex. The results showed that, except for PDB ID: 4I8W, which had a positive $E_{ele}$, all complexes exhibited negative $E_{vdw}$ and $E_{net}$ values (Table A.I), indicating stable protein-ligand interactions.

To determine whether virtual screening requires multiple protein conformations, we computed $E_{net}$ using each complex's protein conformation paired with the ligand conformations from the other 78 complexes. Only 18.1% of ligands exhibited negative $E_{net}$, whereas 70.9% had $E_{net}$ values exceeding 100 kcal/mol, indicating poor binding affinity. These findings highlight the necessity of incorporating multiple protein conformations in structure-based virtual screening, as different ligands may require distinct protein conformations for optimal binding. Since the ideal protein conformation for each ligand is unknown a priori, generating diverse protein conformations is essential. Likewise, considering multiple ligand conformations improves docking and screening accuracy.

**Generation of Protein Conformations via MD Simulations.** Since the native protein conformation in a protein-ligand complex is typically unknown, MD simulations were performed using the apo structure (PDB ID: 3IXO) as the initial conformation without any bound ligand. The protein was solvated and subjected to MD simulations, generating $10^5$ distinct protein conformations.

These conformations were then used to compute protein-ligand binding energies, selecting the conformation with the lowest $E_{net}$ for comparison with experimentally resolved protein-ligand complex structures. The results showed that MD-generated conformations were less favorable than experimentally determined complex structures. Specifically, the number of cases where $E_{ele}$, $E_{vdw}$, and $E_{net}$ were negative decreased to 58, 75, and 74, respectively (Table A.I).

In the original 79 protein-ligand complexes, after filtering out cases with positive binding energies, the average values for $E_{ele}$, $E_{vdw}$, and $E_{net}$ were -78.6, -63.5, and -141.1 kcal/mol, respectively. In contrast, after filtering out positive values for the PDB ID: 3IXO trajectory, the average $E_{ele}$, $E_{vdw}$, and $E_{net}$ were -31.3, -33.9, and -59.6 kcal/mol, respectively. These results suggest that experimentally determined protein-ligand complex structures provide a more favorable binding environment than those generated solely through MD simulations of the apoprotein.

**Generation of Ligand Conformations via MD Simulations.** MD simulations used ligand structures extracted from protein-ligand complexes as initial structures to generate ligand conformations. Each ligand was solvated and subjected to MD simulations, yielding $10^4$ distinct ligand conformations.

To evaluate binding affinity, binding energies were computed for each of the $10^4$ ligand conformations using the protein conformations from the original protein-ligand complex structures. The conformation with the lowest $E_{net}$ was selected and compared to experimentally resolved protein-ligand complex structures. The results indicated that ligand conformations generated from MD simulations were less favorable than those from experimental complex structures. Specifically, the cases where $E_{ele}$, $E_{vdw}$, and $E_{net}$ were negative decreased to 75, 53, and 60, respectively (Table A.I).

After filtering out cases with positive binding energies, the average values of $E_{ele}$, $E_{vdw}$, and $E_{net}$ for MD-generated ligand conformations were -62.0, -31.9, and -91.0 kcal/mol, respectively. These findings suggest that experimentally resolved ligand conformations within the protein-ligand complex provide a more favorable binding environment than those obtained through MD simulations of solvated ligands.

**Reduction in the Number of Conformations Using $RMSD_{thr}$.** Applying $RMSD_{thr}$ effectively reduces the number of protein and ligand conformations, optimizing computational efficiency in docking calculations. As shown in Figure 1, the number of retained conformations decreases exponentially as $RMSD_{thr}$ increases.

For protein structures, starting with $10^5$ trajectory conformations, applying $RMSD_{thr}$ progressively eliminates structurally similar conformations. At lower thresholds (e.g., $RMSD_{thr} < 0.5$ Å), most conformations are retained, whereas at higher thresholds (e.g., $RMSD_{thr} > 1.5$ Å), only a tiny subset remains.

A similar filtering process was applied to ligand structures, where each ligand trajectory consisted of $10^4$ frames. The number of retained conformations decreased with increasing $RMSD_{thr}$, with the final count representing the average across 79 ligands, demonstrating a consistent reduction trend across different ligands.

These results highlight the exponential impact of $RMSD_{thr}$ on reducing redundant conformations, ensuring that structurally diverse conformations are retained while minimizing computational cost.

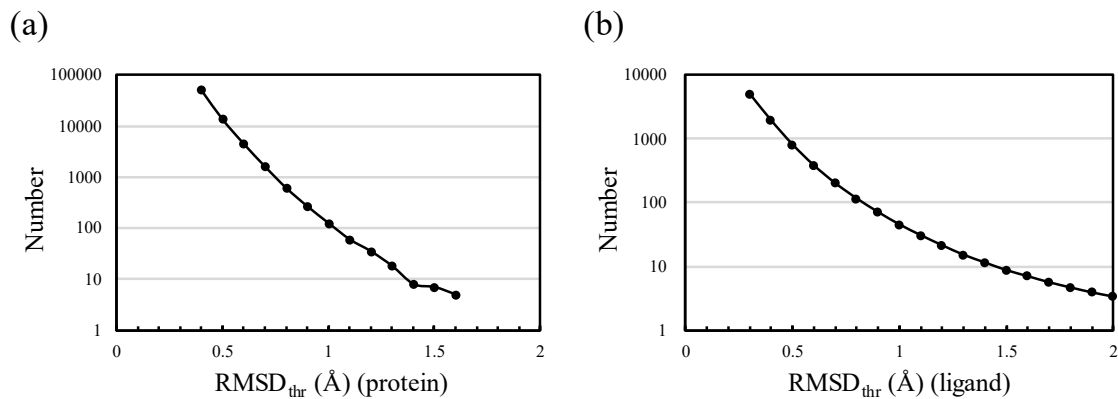

**Figure 1. Reduction in Protein and Ligand Conformations Using RMSD-Based Clustering.** The number of retained conformations is shown as a function of RMSD$_{thr}$, with the vertical axis on a logarithmic scale, illustrating an exponential decrease as RMSD$_{thr}$ increases. (a) Protein conformations were obtained from $10^5$ trajectory frames, with binding pocket residues superimposed and conformations with RMSD < RMSD$_{thr}$ removed. (b) Ligand conformations, where each of the 79 ligand trajectories was individually clustered, and the average number of retained conformations across 79 ligands was computed after superimposing non-hydrogen atoms and removing conformations with RMSD < RMSD$_{thr}$.

**Effect of Clustered Protein and Ligand Conformations on $E_{net}$.** Using clustered protein and ligand conformations, $E_{net}$ was computed to evaluate the impact of structural clustering on protein-ligand interaction energy calculations, as shown in Tables A.II and A.III. The results indicate that as RMSD$_{thr}$ increases, the number of retained protein conformations decreases, leading to variations in $E_{net}$ values.

Figure 2(a) shows that with clustered protein conformations, the proportion of complexes with $E_{net} < 0$ declines as RMSD$_{thr}$ increases, suggesting that fewer favorable binding energies are observed with increasing protein structural diversity. Similarly, Figure 2(b) demonstrates that clustering ligand conformations results in a decreasing proportion of favorable binding energies, highlighting the significant influence of ligand flexibility on interaction energy calculations. Figure 2(c) further examines the combined effect of protein and ligand clustering, revealing that a higher RMSD$_{thr}$ for either protein or ligand clustering tends to reduce the proportion of complexes with $E_{net} < 0$.

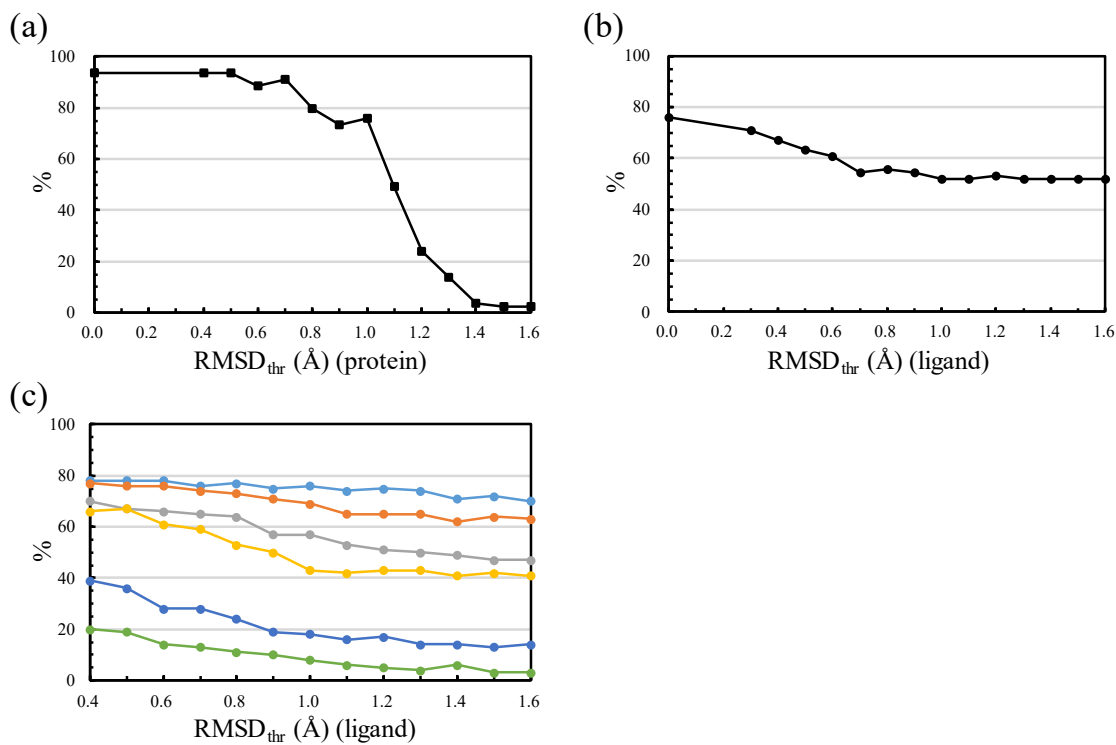

**Figure 2. Effect of Protein and Ligand Clustering on $E_{net}$.** This figure illustrates how protein and ligand conformational clustering influences the proportion of cases with favorable binding energies ($E_{net} < 0$) as a function of $RMSD_{thr}$. (a) The percentage of protein conformations with $E_{net} < 0$ decreases as $RMSD_{thr}$ increases, suggesting reduced binding affinity. (b) A similar decreasing trend is observed for ligand conformations, highlighting the impact of clustering on ligand binding. (c) The combined effect of protein and ligand clustering, where each curve represents a different protein $RMSD_{thr}$: 0.4 (light blue), 0.6 (orange), 0.8 (gray), 1.0 (yellow), 1.2 (blue), and 1.4 (green).

## Discussions

**Protein Binding Pocket Specificity to Ligands.** For the 79 protein-ligand complexes, $E_{net}$ between each protein and its native ligand was consistently negative, indicating favorable interactions. However, when the same protein conformation was used to compute $E_{net}$ for the other 78 ligands, only 18.1% exhibited negative binding energies, while 70.9% had $E_{net} > 100$ kcal/mol, suggesting poor binding affinity.

These findings confirm that incorporating multiple protein conformations in docking simulations is essential. Failing to account for protein flexibility could exclude up to four-fifths of potential inhibitors, significantly compromising the accuracy of virtual screening and drug discovery efforts.

**Generating Protein Conformations via MD Simulations.** Including a ligand in the system is a key consideration in generating protein conformations via MD simulations. If a ligand is present, the protein's conformational space may become restricted, potentially making the generated structures incompatible with binding other ligands. Conversely, without a ligand, the binding pocket interacts with water molecules, which are small, mobile, and highly polar—properties that differ significantly from those of ligands. As a result, longer MD simulations are required to sample the relevant conformational space adequately. In some cases, enhanced sampling techniques may be necessary to overcome energy barriers within a shorter simulation time.

In this study, the apo structure (ligand-free) was used as the initial conformation, and a 100-nanosecond MD simulation was performed, generating $10^5$ trajectory frames. $E_{net}$ calculations between these $10^5$ protein conformations and 79 ligands showed that 74 ligands exhibited $E_{net} < 0$, indicating favorable binding interactions. Extending the MD simulation time and increasing conformational sampling reduced $E_{net}$ values and increased the number of ligands with $E_{net} < 0$ (data not shown). These findings suggest that more extended simulations and enhanced sampling techniques can improve the accuracy of protein conformational ensembles for docking studies.

**Generating Ligand Conformations via MD Simulations.** Ligand conformations were generated using MD simulations, with initial structures extracted from protein-ligand complexes. Since the MD environment consists of water molecules, the sampled ligand conformations favor solvation, as water is highly polar and stabilizes conformations that minimize solvent-exposed potential energy. However, ligands can undergo conformational adjustments upon binding to a protein to better fit within the binding pocket.

This study used ligand conformations from protein-ligand complexes as initial structures and 2-nanosecond MD simulations were performed, collecting $10^4$ trajectory frames per ligand. $E_{net}$ calculations revealed that 60 ligands exhibited $E_{net} < 0$, indicating favorable binding interactions. If ideal ligand conformations were used as initial structures instead, significantly longer simulations and more extensive trajectory sampling would be required to obtain ligand conformations suitable for protein binding (data not shown).

These findings suggest that aqueous-environment-sampled ligand conformations may differ from those needed for optimal protein binding. This reinforces the importance of using protein-bound ligand conformations or enhanced sampling techniques in docking studies.

**Reducing the Impact of Protein and Ligand Conformational Redundancy.** Since the MD simulations in this study were conducted in an aqueous environment, the sampled protein and ligand conformations favored low-energy water states. While higher-energy conformations were occasionally sampled, they occurred less frequently, leading to overrepresenting structurally similar low-energy conformations.

An RMSD-based clustering approach was applied to eliminate redundancy, using an RMSD threshold to filter out structurally similar states. Increasing the RMSD threshold progressively reduced the number of retained conformations by merging similar structures into representative clusters. However, as the number of conformations decreased, $E_{net}$ tended to increase, reflecting the diminished availability of low-energy conformations optimized for binding.

These findings highlight the trade-off between conformational diversity and energy accuracy, emphasizing the importance of selecting an optimal RMSD threshold to balance computational efficiency and binding energy evaluation.

**Conclusion**

SBVS relies on docking to predict protein-ligand interactions. Rigid docking, which assumes a fixed protein conformation, is often employed to reduce computational costs. However, our findings indicate that rigid docking could exclude up to 80% of potential ligands, highlighting the necessity of incorporating multiple protein and ligand conformations to improve docking accuracy. MD simulations were performed to generate diverse protein and ligand conformations to address this limitation. Using the apoprotein (without a ligand) as the initial structure, MD simulations produced $10^5$ distinct protein conformations. Similarly, ligand conformations were generated from MD simulations, using ligand structures from protein-ligand complexes as the starting point, yielding $10^4$ ligand conformations—all contributing to viable binding interactions. To further optimize computational efficiency, RMSD-based clustering was applied to eliminate redundant conformations while preserving structural diversity. However, increasing the RMSD threshold reduced the number of retained conformations, increasing $E_{net}$ values. These findings emphasize the importance of flexible docking and appropriate conformational clustering in virtual screening, ensuring a balance between computational feasibility and accurate ligand-binding predictions.

**Competing interests**

The authors declare no competing interests.

**Funding**

The author received no financial support for the research, authorship, and/or publication of this article.

**Data and Software Availability**

The data supporting the findings of this study are openly available on GitHub at the following URL: https://github.com/peikunyang/09_MD_Clutser

# Leveraging Conformational Diversity for Enhanced Structure-Based Virtual Screening: Insights from Molecular Dynamics Simulations of HIV-1 Protease–Ligand Complexes

Pei-Kun Yang

E-mail: peikun@isu.edu.tw

Table A.I present the computed binding energies for protein-ligand interactions under different conditions. Self-Ligand refers to calculations using the Pro-Lig complex from RCSB, where $E_{ele}$, $E_{vdw}$, and $E_{net}$ were computed for residues within the binding pocket. Cross-Ligand calculations used the protein structure from a given PDB ID with the other 78 ligand structures, reporting the proportion of cases where $E_{net} < 0$ and $E_{net} > 100$ kcal/mol. MD_Pro used MD trajectories from PDB ID: 3IXO, selecting the trajectory frame with the lowest $E_{net}$ when paired with the Pro-Lig complex ligand. MD_Lig involved MD simulations of ligands from the Pro-Lig complex, computing binding energy with the Pro-Lig complex protein and selecting the frame with the lowest $E_{net}$. The unit for $E_{ele}$, $E_{vdw}$, and $E_{net}$ is kcal/mol.

| | Pro-Lig complex | | | | | MD_Pro | | | MD_Lig | | |
|---|---|---|---|---|---|---|---|---|---|---|---|
| | Self-Ligand | | | Cross-Ligand | | | | | | | |
| PDB ID | $E_{net}$ | $E_{ele}$ | $E_{vdw}$ | <0 (%) | >100 (%) | $E_{net}$ | $E_{ele}$ | $E_{vdw}$ | $E_{net}$ | $E_{ele}$ | $E_{vdw}$ |
| 1ajv | -104.9 | -40.3 | -64.6 | 5.1 | 84.6 | -54.9 | -20.2 | -34.8 | -84.2 | -36.4 | -47.8 |
| 1ajx | -106.3 | -45.2 | -61.1 | 19.2 | 68.0 | -63.6 | -28.0 | -35.6 | -70.4 | -36.9 | -33.5 |
| 1c70 | -247.1 | -175.4 | -71.7 | 12.8 | 76.9 | 8.8 | -92.0 | 100.9 | -208.4 | -151.9 | -56.5 |
| 1d4h | -141.2 | -74.1 | -67.1 | 16.7 | 74.4 | -39.4 | -0.7 | -38.7 | -36.3 | -11.5 | -24.8 |
| 1d4i | -145.2 | -74.0 | -71.3 | 18.0 | 74.4 | -38.6 | -2.0 | -36.6 | 127.2 | 4.8 | 122.4 |
| 1d4j | -122.8 | -58.4 | -64.4 | 23.1 | 71.8 | -41.6 | -11.7 | -29.9 | -60.0 | -37.5 | -22.5 |
| 1dif | -129.7 | -43.9 | -85.8 | 30.8 | 53.9 | -57.6 | -10.9 | -46.7 | -45.8 | -45.0 | -0.8 |
| 1ebw | -112.9 | -42.4 | -70.4 | 20.5 | 74.4 | -27.9 | 5.3 | -33.3 | 15.2 | -39.4 | 54.6 |
| 1ebz | -119.9 | -50.9 | -69.0 | 30.8 | 60.3 | -48.2 | -13.2 | -35.0 | -4.1 | -29.7 | 25.7 |
| 1ec0 | -150.1 | -76.9 | -73.2 | 23.1 | 68.0 | -47.6 | -3.9 | -43.8 | -47.8 | -40.5 | -7.3 |
| 1ec1 | -115.0 | -38.1 | -77.0 | 10.3 | 74.4 | -45.8 | -6.0 | -39.8 | 17.3 | -13.9 | 31.3 |
| 1ec2 | -122.9 | -47.2 | -75.7 | 28.2 | 52.6 | -44.6 | -6.9 | -37.7 | -46.8 | -32.5 | -14.2 |
| 1ec3 | -127.9 | -56.8 | -71.1 | 29.5 | 57.7 | -44.3 | -5.5 | -38.9 | -64.9 | -39.5 | -25.4 |
| 1fqx | -150.1 | -86.7 | -63.4 | 5.1 | 78.2 | -11.7 | 31.3 | -43.0 | 110.9 | -79.0 | 189.9 |

| | | | | | | | | | | |
|---|---|---|---|---|---|---|---|---|---|---|
| 1g2k | -106.6 | -41.6 | -65.0 | 10.3 | 82.1 | -57.0 | -18.5 | -38.5 | -77.7 | -28.3 | -49.4 |
| 1g35 | -103.5 | -46.6 | -56.8 | 19.2 | 68.0 | -58.8 | -20.0 | -38.8 | -29.3 | -36.7 | 7.4 |
| 1gno | -68.1 | -34.6 | -33.5 | 52.6 | 30.8 | 48.3 | 0.6 | 47.7 | -31.3 | -15.4 | -15.9 |
| 1hbv | -211.2 | -149.1 | -62.2 | 3.9 | 93.6 | -168.6 | -146.6 | -22.1 | 63.8 | -122.4 | 186.2 |
| 1hih | -118.7 | -55.7 | -63.0 | 9.0 | 84.6 | -38.0 | -0.6 | -37.4 | -87.0 | -38.8 | -48.2 |
| 1hos | -107.3 | -41.0 | -66.3 | 35.9 | 53.9 | -52.5 | -11.9 | -40.6 | 282.3 | -6.8 | 289.1 |
| 1hps | -69.3 | -30.7 | -38.6 | 59.0 | 34.6 | 203.2 | 3.6 | 199.6 | -3.4 | 8.3 | -11.7 |
| 1hpv | -86.1 | -30.2 | -55.9 | 5.1 | 84.6 | -46.9 | -17.2 | -29.7 | -62.5 | -23.4 | -39.1 |
| 1hpx | -111.9 | -47.0 | -65.0 | 15.4 | 69.2 | -42.7 | -7.3 | -35.4 | -26.4 | -22.8 | -3.5 |
| 1hsg | -239.0 | -171.2 | -67.8 | 12.8 | 70.5 | -146.0 | -119.9 | -26.1 | -207.0 | -160.4 | -46.6 |
| 1htf | -214.6 | -161.3 | -53.4 | 12.8 | 73.1 | -91.7 | -64.1 | -27.6 | -176.0 | -150.9 | -25.1 |
| 1htg | -105.4 | -82.1 | -23.3 | 69.2 | 19.2 | 294.2 | -49.8 | 344.0 | 19.0 | -62.1 | 81.1 |
| 1hvi | -136.3 | -60.8 | -75.6 | 34.6 | 55.1 | -36.8 | 11.9 | -48.7 | -32.1 | -18.5 | -13.6 |
| 1hvj | -310.6 | -233.3 | -77.3 | 28.2 | 59.0 | -176.6 | -147.1 | -29.5 | -79.8 | -217.8 | 138.0 |
| 1hvk | -129.4 | -49.3 | -80.0 | 29.5 | 59.0 | -33.9 | 6.4 | -40.3 | 130.2 | -0.9 | 131.1 |
| 1hvl | -122.5 | -48.8 | -73.6 | 33.3 | 60.3 | -28.0 | 0.3 | -28.2 | 30.7 | 14.1 | 16.6 |
| 1iiq | -191.0 | -124.8 | -66.2 | 21.8 | 57.7 | -104.3 | -70.8 | -33.6 | -142.5 | -97.0 | -45.5 |
| 1m0b | -115.2 | -47.6 | -67.6 | 14.1 | 71.8 | -19.7 | 13.4 | -33.1 | -62.2 | -43.7 | -18.5 |
| 1mui | -119.2 | -52.9 | -66.3 | 33.3 | 53.9 | -42.3 | -11.1 | -31.2 | 22.1 | -39.7 | 61.8 |
| 1npa | -240.5 | -175.0 | -65.6 | 20.5 | 64.1 | -162.2 | -134.4 | -27.8 | -122.7 | -155.9 | 33.2 |
| 1npv | -95.3 | -30.7 | -64.6 | 11.5 | 80.8 | -40.9 | -2.9 | -38.0 | -30.3 | -1.1 | -29.2 |
| 1npw | -93.8 | -25.1 | -68.6 | 28.2 | 59.0 | -42.2 | -1.1 | -41.0 | 42.0 | -5.4 | 47.4 |
| 1ohr | -250.9 | -187.3 | -63.6 | 23.1 | 61.5 | -112.7 | -80.5 | -32.2 | -173.7 | -133.0 | -40.7 |
| 1t7k | -108.7 | -41.3 | -67.5 | 12.8 | 82.1 | -55.9 | -21.9 | -34.0 | -78.7 | -34.9 | -43.9 |
| 1wbk | -112.8 | -44.4 | -68.4 | 18.0 | 74.4 | -41.3 | -2.9 | -38.4 | 10.4 | -6.3 | 16.7 |
| 1wbm | -137.5 | -71.0 | -66.5 | 16.7 | 68.0 | -36.6 | -12.9 | -23.7 | 52.8 | -20.3 | 73.1 |
| 1xl2 | -153.1 | -104.5 | -48.6 | 2.6 | 94.9 | -41.9 | -36.5 | -5.4 | -77.9 | -106.4 | 28.5 |
| 1xl5 | -112.8 | -50.9 | -61.9 | 11.5 | 70.5 | -40.8 | -4.3 | -36.6 | -78.8 | -37.0 | -41.8 |
| 1yt9 | -112.8 | -50.5 | -62.3 | 14.1 | 78.2 | -43.0 | -9.6 | -33.4 | -24.8 | -2.3 | -22.5 |
| 1zsf | -167.2 | -105.4 | -61.8 | 28.2 | 53.9 | -20.0 | 17.1 | -37.1 | 10.4 | -95.5 | 105.8 |
| 2a4f | -137.3 | -67.2 | -70.1 | 19.2 | 70.5 | -35.9 | 4.9 | -40.8 | -39.2 | -36.5 | -2.6 |
| 2aqu | -119.0 | -47.3 | -71.7 | 24.4 | 50.0 | -54.2 | -14.2 | -40.0 | 30.4 | -22.9 | 53.2 |
| 2bb9 | -104.0 | -35.9 | -68.1 | 25.6 | 61.5 | -38.4 | -14.2 | -24.2 | 14.8 | -23.5 | 38.3 |
| 2bbb | -105.9 | -36.3 | -69.6 | 10.3 | 78.2 | -46.3 | -3.0 | -43.3 | -21.7 | -6.6 | -15.1 |
| 2bpv | -168.8 | -102.9 | -65.9 | 11.5 | 78.2 | -110.0 | -78.2 | -31.9 | -90.4 | -51.8 | -38.6 |
| 2bpy | -211.8 | -145.3 | -66.5 | 9.0 | 80.8 | -154.4 | -113.7 | -40.7 | -173.8 | -121.2 | -52.6 |
| 2bqv | -103.1 | -33.5 | -69.6 | 14.1 | 66.7 | -52.1 | -12.1 | -40.1 | -30.4 | -10.1 | -20.3 |
| 2cej | -110.4 | -37.8 | -72.6 | 6.4 | 89.7 | -25.8 | -3.2 | -22.6 | 17.2 | -6.7 | 23.9 |
| 2cem | -114.4 | -37.3 | -77.1 | 28.2 | 62.8 | -31.1 | -19.3 | -11.7 | 18.7 | -19.2 | 37.9 |
| 2cen | -118.6 | -43.0 | -75.7 | 18.0 | 71.8 | -45.3 | -19.9 | -25.4 | 42.1 | -22.8 | 64.9 |
| 2pqz | -161.8 | -100.5 | -61.3 | 10.3 | 85.9 | -82.0 | -47.5 | -34.6 | -150.9 | -99.6 | -51.3 |
| 2pwc | -267.7 | -207.9 | -59.8 | 9.0 | 89.7 | -86.7 | -51.7 | -35.0 | -244.1 | -197.1 | -46.9 |
| 2pwr | -268.2 | -204.1 | -64.1 | 7.7 | 87.2 | -72.5 | -31.9 | -40.6 | -246.3 | -200.7 | -45.6 |
| 2qnn | -274.0 | -205.9 | -68.1 | 14.1 | 76.9 | -67.1 | -38.4 | -28.8 | -245.1 | -202.2 | -42.9 |
| 2qnp | -280.9 | -219.2 | -61.7 | 10.3 | 82.1 | -80.4 | -45.2 | -35.3 | -266.4 | -209.4 | -56.9 |
| 2qnq | -259.2 | -202.8 | -56.4 | 14.1 | 75.6 | -84.3 | -49.5 | -34.8 | -248.3 | -200.8 | -47.4 |
| 2upj | -102.8 | -40.2 | -62.6 | 1.3 | 94.9 | -84.1 | -37.9 | -46.2 | -51.9 | -32.2 | -19.7 |
| 2uxz | -108.6 | -36.2 | -72.4 | 9.0 | 88.5 | -47.5 | -9.4 | -38.2 | -15.4 | -20.5 | 5.1 |
| 2uy0 | -124.5 | -49.6 | -74.8 | 9.0 | 79.5 | -54.1 | -14.9 | -39.2 | -36.2 | -21.6 | -14.6 |
| 2wkz | -120.9 | -42.9 | -78.1 | 15.4 | 71.8 | -57.1 | -14.2 | -43.0 | -48.1 | -26.3 | -21.8 |
| 3bgb | -217.8 | -163.2 | -54.7 | 16.7 | 65.4 | -78.4 | -48.5 | -29.9 | -192.4 | -169.7 | -22.6 |
| 3bgc | -149.8 | -91.9 | -57.8 | 9.0 | 80.8 | -69.7 | -51.6 | -18.1 | -112.8 | -83.0 | -29.8 |
| 3bhe | -143.0 | -109.7 | -33.2 | 2.6 | 94.9 | -57.9 | -39.3 | -18.6 | -118.9 | -110.3 | -8.6 |
| 3ckt | -177.4 | -119.6 | -57.8 | 14.1 | 76.9 | -72.4 | -44.3 | -28.1 | -158.2 | -102.2 | -56.0 |
| 3t11 | -41.8 | -3.1 | -38.6 | 14.1 | 76.9 | -20.1 | -1.4 | -18.8 | -26.7 | -4.4 | -22.3 |
| 4hla | -96.9 | -44.2 | -52.7 | 15.4 | 69.2 | -48.5 | -19.0 | -29.5 | -42.7 | -18.1 | -24.5 |
| 4i8w | -47.7 | 3.3 | -51.0 | 12.8 | 69.2 | 30.3 | 61.4 | -31.0 | -8.9 | 26.5 | -35.4 |
| 4i8z | -97.5 | -40.1 | -57.4 | 11.5 | 74.4 | -55.0 | -14.7 | -40.3 | -26.9 | -48.9 | 22.0 |
| 5tyr | -123.8 | -62.0 | -61.8 | 15.4 | 79.5 | -49.8 | -13.9 | -36.0 | -90.0 | -47.5 | -42.5 |
| 5tys | -108.1 | -51.5 | -56.6 | 10.3 | 69.2 | -58.3 | -22.8 | -35.5 | -60.5 | -42.0 | -18.5 |
| 6d0d | -110.3 | -53.5 | -56.8 | 11.5 | 80.8 | -40.9 | -7.2 | -33.7 | -97.9 | -45.4 | -52.5 |
| 6d0e | -113.8 | -58.5 | -55.3 | 18.0 | 74.4 | -38.8 | -4.6 | -34.2 | -83.3 | -49.9 | -33.4 |
| 6uwb | -127.5 | -69.1 | -58.4 | 29.5 | 53.9 | -49.2 | -10.5 | -38.7 | -107.5 | -63.5 | -44.0 |
| 6uwc | -106.3 | -43.0 | -63.4 | 18.0 | 71.8 | -66.3 | -26.6 | -39.7 | -81.4 | -31.8 | -49.6 |
| 7upj | -74.0 | -23.7 | -50.3 | 2.6 | 83.3 | -49.1 | -15.9 | -33.2 | -70.7 | -22.8 | -47.9 |

Table A.II: Computed $E_{net}$ for Protein-Ligand Interactions. This table presents $E_{net}$ values for protein-ligand interactions using clustered protein conformations, considering only electrostatic and van der Waals interactions between residues within the binding pocket and the ligand. The unit is kcal/mol.

| PDB ID | Complex | RMSD$_{thr}$ | | | | | | | | |
|---|---|---|---|---|---|---|---|---|---|---|
| | | 0 | 0.4 | 0.5 | 0.6 | 0.7 | 0.8 | 0.9 | 1 | 1.1 |
| 1ajv | -104.9 | -54.9 | -50.6 | -45.5 | -43.5 | -40.9 | -33.0 | -36.4 | -37.2 | -7.5 |
| 1ajx | -106.3 | -63.6 | -63.6 | -58.5 | -58.8 | -58.8 | -47.5 | -45.1 | -40.2 | -34.1 |
| 1c70 | -247.1 | 8.8 | 131.4 | 131.4 | 255.1 | 317.8 | 625.7 | 1867.0 | 3213.3 | 70.6 |
| 1d4h | -141.2 | -39.4 | -39.4 | -39.4 | -18.7 | -39.4 | -5.0 | 10.8 | -18.7 | 62.9 |
| 1d4i | -145.2 | -38.6 | -38.6 | -38.6 | -32.3 | -38.6 | -1.8 | 38.2 | -2.4 | 61.6 |
| 1d4j | -122.8 | -41.6 | -41.6 | -41.4 | 32.6 | -41.4 | -5.5 | 61.3 | 25.0 | 174.7 |
| 1dif | -129.7 | -57.6 | -57.4 | -57.4 | -57.4 | -29.2 | -14.3 | -24.5 | -45.7 | 80.6 |
| 1ebw | -112.9 | -27.9 | -27.9 | -27.9 | -21.4 | -27.9 | -6.4 | 17.4 | -3.9 | 58.5 |
| 1ebz | -119.9 | -48.2 | -48.2 | -41.2 | -37.7 | -41.2 | -24.6 | -16.4 | -31.5 | 12.0 |
| 1ec0 | -150.1 | -47.6 | -47.6 | -47.6 | -40.2 | -47.6 | -11.3 | -2.1 | -9.7 | 19.7 |
| 1ec1 | -115.0 | -45.8 | -45.8 | -39.2 | -34.1 | -39.2 | 1.0 | -45.8 | -32.5 | -20.1 |
| 1ec2 | -122.9 | -44.6 | -44.6 | -39.7 | -35.0 | -39.7 | -6.2 | -44.6 | -33.5 | -18.5 |
| 1ec3 | -127.9 | -44.3 | -44.3 | -39.6 | -37.8 | -22.9 | -10.7 | -28.3 | -21.7 | -26.6 |
| 1fqx | -150.1 | -11.7 | -11.7 | -1.9 | 0.1 | -0.4 | 14.4 | -1.9 | 12.5 | 14.4 |
| 1g2k | -106.7 | -57.0 | -57.0 | -44.6 | -44.6 | -38.5 | -40.3 | -36.5 | -39.6 | -3.6 |
| 1g35 | -103.5 | -58.8 | -50.0 | -41.8 | -40.2 | -38.2 | -37.8 | -28.1 | -32.8 | 32.1 |
| 1gno | -68.1 | 48.3 | 51.0 | 51.0 | 69.7 | 89.0 | 770.7 | 1126.7 | 563.3 | 2511.0 |
| 1hbv | -211.2 | -168.6 | -168.6 | -167.1 | -155.6 | -154.0 | -167.1 | -144.5 | -140.7 | -137.5 |
| 1hih | -118.7 | -38.0 | -38.0 | -31.5 | -28.6 | -24.1 | 4.4 | -7.9 | -14.4 | 16.8 |
| 1hos | -107.3 | -52.5 | -51.6 | -46.2 | -40.5 | -38.4 | -37.3 | -37.6 | -37.8 | -11.7 |
| 1hps | -69.3 | 203.2 | 203.2 | 354.5 | 354.5 | 1178.3 | 296.7 | 370.6 | 1145.9 | 1339.5 |
| 1hpv | -86.1 | -46.9 | -46.9 | -44.0 | -41.4 | -35.6 | -40.5 | -32.6 | -31.8 | -35.0 |
| 1hpx | -111.9 | -42.7 | -42.7 | -42.7 | -26.9 | -36.4 | -25.3 | -39.7 | 30.0 | -0.2 |
| 1hsg | -239.0 | -146.0 | -142.5 | -146.0 | -127.9 | -142.5 | -114.3 | -102.7 | -127.9 | -92.5 |
| 1htf | -214.6 | -91.7 | -91.7 | -89.4 | -79.9 | -74.0 | -64.4 | -73.0 | -74.6 | -42.3 |
| 1htg | -105.4 | 294.2 | 294.2 | 653.7 | 1780.7 | 294.2 | 5762.5 | 5762.5 | 7144.2 | 1004.2 |
| 1hvi | -136.4 | -36.8 | -36.8 | -27.9 | -33.8 | -23.1 | -14.8 | -8.5 | -14.7 | 49.5 |

| ID | | | | | | | | | |
|---|---|---|---|---|---|---|---|---|---|
| 1hvj | -310.6 | -176.6 | -176.6 | -176.6 | -161.2 | -153.9 | -176.6 | -146.0 | -167.0 | -102.1 |
| 1hvk | -129.4 | -33.9 | -30.7 | -30.7 | -16.0 | -15.6 | 1.0 | -13.4 | -26.2 | 165.1 |
| 1hvl | -122.5 | -28.0 | -24.0 | -20.1 | -4.4 | -7.0 | 93.2 | -11.3 | -9.4 | 106.0 |
| 1iiq | -191.0 | -104.3 | -104.3 | -90.8 | -86.5 | -104.3 | -89.0 | -92.7 | -77.8 | -87.5 |
| 1m0b | -115.2 | -19.7 | -19.7 | -8.9 | -5.9 | -7.7 | -8.1 | -8.9 | 9.3 | 3.7 |
| 1mui | -119.2 | -42.3 | -40.1 | -39.6 | -36.1 | -34.5 | -28.4 | -17.6 | -32.8 | -4.8 |
| 1npa | -240.5 | -162.2 | -162.2 | -162.0 | -151.8 | -157.1 | -119.7 | -123.9 | -132.3 | -132.3 |
| 1npv | -95.3 | -40.9 | -40.9 | -34.4 | -22.9 | -11.8 | -14.2 | -1.4 | 43.1 | 80.8 |
| 1npw | -93.8 | -42.2 | -37.9 | -42.2 | -33.9 | -37.9 | -6.8 | -26.1 | -26.0 | -20.4 |
| 1ohr | -250.9 | -112.7 | -105.4 | -98.1 | -105.4 | -90.7 | -84.8 | -95.7 | -73.8 | -78.3 |
| 1t7k | -108.7 | -55.9 | -55.9 | -52.3 | -46.3 | -47.1 | -47.5 | -42.4 | -35.4 | -22.7 |
| 1wbk | -112.8 | -41.3 | -41.3 | -41.3 | -0.2 | -41.3 | -2.5 | 10.6 | 13.4 | 138.0 |
| 1wbm | -137.5 | -36.6 | -36.6 | -29.7 | 11.1 | -29.7 | 15.6 | 59.1 | 162.4 | 284.6 |
| 1xl2 | -153.1 | -41.9 | -41.9 | -36.5 | -36.3 | -36.5 | 270.9 | 4800.6 | 952.5 | 8440.2 |
| 1xl5 | -112.8 | -40.8 | -40.8 | -40.8 | -19.2 | -40.8 | -22.5 | 4.5 | -14.7 | 172.7 |
| 1yt9 | -112.8 | -43.0 | -43.0 | -43.0 | -35.8 | -32.9 | -6.3 | 11.0 | -23.9 | 105.1 |
| 1zsf | -167.2 | -20.0 | -10.5 | -10.5 | -1.4 | 26.0 | 34.1 | 7.5 | -1.4 | 49.5 |
| 2a4f | -137.3 | -35.9 | -35.9 | -30.4 | -21.9 | -32.4 | -13.9 | -31.2 | -20.2 | -1.9 |
| 2aqu | -119.0 | -54.2 | -54.2 | -54.2 | -53.0 | -53.0 | -33.7 | -41.1 | -34.3 | 9.6 |
| 2bb9 | -104.0 | -38.4 | -27.2 | -25.2 | -4.6 | -6.4 | 101.4 | 134.7 | 167.3 | 165.3 |
| 2bbb | -105.9 | -46.3 | -45.2 | -44.4 | -43.2 | -44.4 | -22.1 | -40.2 | -38.3 | -27.7 |
| 2bpv | -168.8 | -110.0 | -109.6 | -104.6 | -104.3 | -104.3 | -99.2 | -96.5 | -100.3 | -90.1 |
| 2bpy | -211.8 | -154.4 | -154.4 | -146.5 | -145.0 | -139.5 | -133.5 | -133.5 | -137.9 | -130.1 |
| 2bqv | -103.1 | -52.1 | -52.1 | -45.0 | -45.3 | -40.7 | -25.0 | -35.9 | -33.4 | -27.5 |
| 2cej | -110.4 | -25.8 | -25.2 | -24.3 | -15.2 | -11.0 | -17.9 | 40.2 | 34.3 | 90.0 |
| 2cem | -114.4 | -31.1 | -28.2 | -31.1 | -19.8 | -19.9 | -1.9 | -1.2 | -13.3 | 114.5 |
| 2cen | -118.6 | -45.3 | -41.7 | -45.3 | -28.2 | -32.1 | -34.3 | -9.3 | -25.6 | 12.9 |
| 2pqz | -161.8 | -82.0 | -78.8 | -70.1 | -71.7 | -68.0 | -64.0 | -65.6 | -66.2 | -52.9 |
| 2pwc | -267.7 | -86.7 | -84.6 | -84.6 | -84.6 | -70.0 | -70.6 | -71.1 | -70.9 | -64.5 |
| 2pwr | -268.2 | -72.5 | -68.9 | -62.9 | -57.5 | -59.4 | -59.6 | -59.8 | -60.2 | -26.1 |
| 2qnn | -274.0 | -67.1 | -67.1 | -58.5 | -40.9 | -40.7 | -46.9 | -45.4 | -44.5 | 43.0 |
| 2qnp | -280.9 | -80.4 | -79.7 | -76.2 | -76.2 | -64.3 | -64.3 | -65.8 | -65.3 | -38.4 |
| 2qnq | -259.2 | -84.3 | -84.3 | -74.4 | -76.7 | -68.2 | -68.2 | -69.1 | -57.8 | -54.7 |

| ID | | | | | | | | | |
|------|--------|-------|-------|-------|-------|-------|-------|-------|-------|
| 2upj | -102.9 | -84.1 | -75.9 | -75.1 | -69.3 | -61.9 | -50.6 | -37.4 | -66.3 | -38.4 |
| 2uxz | -108.6 | -47.5 | -42.6 | -44.1 | -27.9 | -7.2 | -43.8 | 32.6 | 65.5 | 293.7 |
| 2uy0 | -124.5 | -54.1 | -46.4 | -54.1 | -23.8 | -28.2 | -38.6 | 6.2 | 18.2 | 257.2 |
| 2wkz | -120.9 | -57.1 | -57.1 | -43.8 | -36.8 | -42.0 | -33.5 | -40.3 | 13.4 | 20.0 |
| 3bgb | -217.8 | -78.4 | -78.4 | -69.2 | -60.2 | -58.1 | -55.9 | -41.6 | -50.8 | 37.1 |
| 3bgc | -149.8 | -69.7 | -69.7 | -54.0 | 1.2 | 7.3 | 101.0 | 116.6 | -25.4 | 466.7 |
| 3bhe | -143.0 | -57.9 | -57.8 | -51.2 | -46.9 | -51.2 | -12.5 | -33.0 | -23.9 | -30.7 |
| 3ckt | -177.4 | -72.4 | -72.4 | -68.7 | -53.4 | -55.1 | -43.1 | -55.9 | -45.4 | -12.2 |
| 3t11 | -41.8 | -20.1 | -20.1 | -9.1 | -16.9 | -4.0 | 56.2 | 124.4 | 58.0 | 334.7 |
| 4hla | -96.9 | -48.5 | -44.0 | -41.6 | -39.4 | -33.4 | -26.3 | -27.0 | -22.4 | -30.7 |
| 4i8w | -47.7 | 30.3 | 30.3 | 41.1 | 38.1 | 30.3 | 61.2 | 46.6 | 59.2 | 66.3 |
| 4i8z | -97.5 | -55.0 | -55.0 | -49.9 | -44.6 | -41.9 | -55.0 | -38.7 | -29.1 | -29.0 |
| 5tyr | -123.8 | -49.8 | -44.8 | -42.1 | -40.5 | -44.8 | -20.7 | -17.3 | -32.6 | 136.0 |
| 5tys | -108.1 | -58.3 | -58.3 | -46.1 | -37.8 | -40.5 | -18.9 | -39.3 | -38.7 | 4.7 |
| 6d0d | -110.3 | -40.9 | -40.9 | -39.0 | -35.7 | -37.6 | -35.7 | -32.8 | -23.9 | -26.3 |
| 6d0e | -113.8 | -38.8 | -38.8 | -37.0 | -34.7 | -37.0 | -29.3 | -23.3 | -23.9 | -6.1 |
| 6uwb | -127.5 | -49.2 | -49.2 | -43.2 | -42.8 | -38.5 | -24.7 | -43.2 | -25.1 | -16.9 |
| 6uwc | -106.3 | -66.3 | -66.3 | -55.4 | -57.3 | -52.9 | -66.3 | -34.2 | -46.1 | -34.6 |
| 7upj | -74.0 | -49.1 | -49.1 | -48.1 | -40.6 | -42.7 | -33.8 | -27.0 | -28.2 | -14.5 |

Table A.III: Computed $E_{net}$ for Protein-Ligand Interactions. This table presents $E_{net}$ values for protein-ligand interactions using clustered ligand conformations, considering only electrostatic and van der Waals interactions between residues within the binding pocket and the ligand. The unit is kcal/mol.

| PDB ID | Complex | RMSD$_{thr}$ | | | | | | | | |
|---|---|---|---|---|---|---|---|---|---|---|
| | | 0 | 0.3 | 0.4 | 0.5 | 0.6 | 0.7 | 0.8 | 0.9 | 1 |
| 1ajv | -104.9 | -84.2 | -62.3 | -41.3 | -37.0 | -37.0 | -37.0 | -37.0 | -37.0 | -37.0 |
| 1ajx | -106.3 | -70.4 | -70.4 | -70.4 | -70.4 | -70.4 | -70.4 | -70.4 | -70.4 | -70.4 |
| 1c70 | -247.1 | -208.4 | -208.4 | -208.4 | -208.4 | -208.4 | -208.4 | -208.4 | -208.4 | -208.4 |
| 1d4h | -141.2 | -36.3 | -36.3 | -36.3 | 236.9 | 539.7 | 606.7 | 799.7 | 799.7 | 799.7 |
| 1d4i | -145.2 | 127.2 | 128.3 | 160.9 | 160.9 | 160.9 | 160.9 | 160.9 | 160.9 | 160.9 |
| 1d4j | -122.8 | -60.0 | -60.0 | -60.0 | -60.0 | -60.0 | -60.0 | -60.0 | -60.0 | -60.0 |
| 1dif | -129.7 | -45.8 | -45.8 | 8.2 | 8.2 | 8.2 | 8.2 | 8.2 | 8.2 | 8.2 |
| 1ebw | -112.9 | 15.2 | 15.2 | 15.2 | 109.8 | 109.8 | 109.8 | 109.8 | 109.8 | 109.8 |
| 1ebz | -119.9 | -4.1 | 18.0 | -4.1 | 320.3 | 320.3 | 320.3 | 320.3 | 320.3 | 320.3 |
| 1ec0 | -150.1 | -47.8 | -47.8 | -47.8 | -47.8 | -47.8 | -47.8 | -47.8 | -47.8 | -47.8 |
| 1ec1 | -115.0 | 17.3 | 17.3 | 17.3 | 17.3 | 17.3 | 17.3 | 17.3 | 17.3 | 17.3 |
| 1ec2 | -122.9 | -46.8 | -46.8 | -46.8 | -46.8 | -46.8 | -46.8 | -46.8 | -46.8 | -46.8 |
| 1ec3 | -127.9 | -64.9 | -64.9 | -64.9 | -64.9 | -64.9 | -64.9 | -64.9 | -64.9 | -64.9 |
| 1fqx | -150.1 | 110.9 | 110.9 | 294.5 | 294.5 | 450.7 | 2130.0 | 2130.0 | 2130.0 | 2130.0 |
| 1g2k | -106.7 | -77.7 | -77.7 | -77.7 | -26.5 | 251.6 | 208.2 | 131.7 | 316.7 | 316.7 |
| 1g35 | -103.5 | -29.3 | -15.4 | 2.6 | 35.7 | -14.3 | 37.0 | 37.0 | 37.0 | 37.0 |
| 1gno | -68.1 | -31.3 | -31.3 | -31.3 | -17.1 | 11.0 | 11.0 | 25.3 | 25.3 | 25.3 |
| 1hbv | -211.2 | 63.8 | 63.8 | 63.8 | 2060.1 | 486.7 | 4454.2 | 897.6 | 4454.2 | 4454.2 |
| 1hih | -118.7 | -87.0 | -87.0 | -79.2 | -67.3 | -53.8 | -41.1 | -41.1 | -41.1 | -41.1 |
| 1hos | -107.3 | 282.3 | 460.0 | 460.0 | 460.0 | 460.0 | 460.0 | 460.0 | 460.0 | 460.0 |
| 1hps | -69.3 | -3.4 | -2.5 | -2.5 | 4.4 | 43.8 | 228.0 | 61.0 | 519.3 | 78.5 |
| 1hpv | -86.1 | -62.5 | -58.4 | -58.4 | -45.2 | -50.4 | -5.9 | -11.6 | -5.9 | 14.1 |
| 1hpx | -111.9 | -26.4 | -26.4 | -26.4 | -26.4 | -26.4 | -26.4 | -26.4 | -26.4 | -26.4 |
| 1hsg | -239.0 | -207.0 | -207.0 | -156.9 | -156.9 | -156.9 | -156.9 | -156.9 | -156.9 | -156.9 |
| 1htf | -214.6 | -176.0 | -176.0 | -176.0 | -176.0 | -176.0 | -176.0 | -176.0 | -176.0 | -176.0 |
| 1htg | -105.4 | 19.0 | 131.7 | 531.0 | 295.0 | 531.0 | 531.0 | 531.0 | 295.0 | 531.0 |
| 1hvi | -136.4 | -32.1 | -32.1 | 45.9 | 136.6 | -32.1 | 339.6 | 369.4 | 369.4 | 369.4 |

| | | | | | | | | | |
|---|---|---|---|---|---|---|---|---|---|
| 1hvj | -310.6 | -79.8 | -74.4 | -74.4 | -74.4 | -74.4 | -74.4 | -74.4 | -74.4 |
| 1hvk | -129.4 | 130.2 | 130.2 | 130.2 | 130.2 | 130.2 | 130.2 | 130.2 | 130.2 |
| 1hvl | -122.5 | 30.7 | 30.7 | 30.7 | 119.5 | 119.5 | 119.5 | 119.5 | 119.5 |
| 1iiq | -191.0 | -142.5 | -142.5 | -142.5 | -142.5 | -142.5 | -142.5 | -142.5 | -142.5 |
| 1m0b | -115.2 | -62.2 | -57.1 | 5.3 | 106.6 | 106.6 | 106.6 | 106.6 | 106.6 |
| 1mui | -119.2 | 22.1 | 22.1 | 150.9 | 169.2 | 169.2 | 169.2 | 169.2 | 169.2 |
| 1npa | -240.5 | -122.7 | -122.7 | -122.7 | -12.3 | -12.3 | -12.3 | -12.3 | -12.3 |
| 1npv | -95.3 | -30.3 | -16.8 | -16.8 | -16.8 | -16.8 | -16.8 | -16.8 | -16.8 |
| 1npw | -93.8 | 42.0 | 45.0 | 112.6 | 112.6 | 112.6 | 112.6 | 112.6 | 112.6 |
| 1ohr | -250.9 | -173.7 | -173.7 | -173.7 | -173.7 | -173.7 | -173.7 | -173.7 | -173.7 |
| 1t7k | -108.7 | -78.7 | -78.7 | -78.7 | -78.7 | -78.7 | -78.7 | -78.7 | -78.7 |
| 1wbk | -112.8 | 10.4 | 10.4 | 10.4 | 42.7 | 42.7 | 42.7 | 42.7 | 42.7 |
| 1wbm | -137.5 | 52.8 | 52.8 | 52.8 | 52.8 | 52.8 | 52.8 | 52.8 | 52.8 |
| 1xl2 | -153.1 | -77.9 | -74.4 | -65.6 | -22.1 | -17.9 | 38.5 | 38.5 | 38.5 |
| 1xl5 | -112.8 | -78.8 | -78.8 | -72.2 | -72.2 | -72.2 | -72.2 | -72.2 | -72.2 |
| 1yt9 | -112.8 | -24.8 | 31.1 | 31.1 | 51.6 | 51.6 | 51.6 | 51.6 | 51.6 |
| 1zsf | -167.2 | 10.4 | 35.5 | 35.5 | 35.5 | 35.5 | 35.5 | 35.5 | 35.5 |
| 2a4f | -137.3 | -39.2 | -39.2 | -39.2 | -39.2 | -39.2 | -39.2 | -39.2 | -39.2 |
| 2aqu | -119.0 | 30.4 | 30.4 | 98.5 | 362.9 | 362.9 | 393.8 | 393.8 | 393.8 |
| 2bb9 | -104.0 | 14.8 | 14.8 | 14.8 | 14.8 | 14.8 | 14.8 | 14.8 | 14.8 |
| 2bbb | -105.9 | -21.7 | 3.5 | 29.0 | 41.8 | 146.6 | 146.6 | 146.6 | 146.6 |
| 2bpv | -168.8 | -90.4 | -90.4 | -86.8 | -86.8 | -86.8 | -86.8 | -86.8 | -86.8 |
| 2bpy | -211.8 | -173.8 | -173.8 | -173.1 | -173.1 | -173.1 | -173.1 | -173.1 | -173.1 |
| 2bqv | -103.1 | -30.4 | -30.4 | -30.4 | -30.4 | -30.4 | -30.4 | -30.4 | -30.4 |
| 2cej | -110.4 | 17.2 | 17.2 | 494.2 | 881.8 | 330.8 | 881.8 | 881.8 | 881.8 |
| 2cem | -114.4 | 18.7 | 18.7 | 18.7 | 18.7 | 18.7 | 18.7 | 18.7 | 18.7 |
| 2cen | -118.6 | 42.1 | 42.1 | 42.1 | 278.4 | 278.4 | 278.4 | 278.4 | 278.4 |
| 2pqz | -161.8 | -150.9 | -143.5 | -143.2 | -103.9 | -103.9 | -103.9 | -103.9 | -103.9 |
| 2pwc | -267.7 | -244.1 | -244.1 | -244.1 | -244.1 | -244.1 | -244.1 | -244.1 | -244.1 |
| 2pwr | -268.2 | -246.3 | -246.3 | -189.5 | -190.2 | -188.2 | -148.0 | -148.0 | -148.0 |
| 2qnn | -274.0 | -245.1 | -239.5 | -229.1 | -219.9 | -172.3 | -172.3 | -172.3 | -172.3 |
| 2qnp | -280.9 | -266.4 | -266.4 | -266.4 | -266.4 | -266.4 | -266.4 | -266.4 | -266.4 |
| 2qnq | -259.2 | -248.3 | -238.4 | -221.8 | -200.4 | -200.4 | -200.4 | -200.4 | -200.4 |

| ID | | | | | | | | | |
|------|--------|--------|--------|--------|--------|--------|--------|--------|--------|
| 2upj | -102.9 | -51.9  | -51.9  | -26.1  | -26.1  | -26.1  | -26.1  | -26.1  | -26.1  |
| 2uxz | -108.6 | -15.4  | -15.4  | -15.4  | -15.4  | -15.4  | -15.4  | -15.4  | -15.4  |
| 2uy0 | -124.5 | -36.2  | -36.2  | -36.2  | -36.2  | -36.2  | -36.2  | -36.2  | -36.2  |
| 2wkz | -120.9 | -48.1  | -45.1  | -48.1  | -9.7   | 54.3   | 89.0   | 89.0   | 89.0   |
| 3bgb | -217.8 | -192.4 | -192.4 | -174.0 | -191.6 | -170.0 | -172.5 | -179.2 | -170.0 |
| 3bgc | -149.8 | -112.8 | -88.9  | -67.4  | -67.4  | -67.4  | -67.4  | -67.4  | -67.4  |
| 3bhe | -143.0 | -118.9 | -106.9 | -101.4 | -101.4 | -101.4 | -101.4 | -101.4 | -101.4 |
| 3ckt | -177.4 | -158.2 | -147.2 | -147.2 | -147.2 | -143.5 | -143.5 | -143.5 | -143.5 |
| 3t11 | -41.8  | -26.7  | -26.1  | -18.3  | -4.1   | -4.1   | -4.1   | -4.1   | -10.5  |
| 4hla | -96.9  | -42.7  | -42.7  | -15.4  | -11.5  | -6.4   | 79.4   | 64.6   | 117.8  |
| 4i8w | -47.7  | -8.9   | 6.8    | 6.1    | -8.9   | 131.5  | 289.2  | 73.7   | 327.9  |
| 4i8z | -97.5  | -26.9  | -14.1  | -10.5  | 13.2   | 13.2   | 107.1  | 591.2  | 168.6  |
| 5tyr | -123.8 | -90.0  | -84.6  | -80.6  | -80.6  | -80.6  | -80.6  | -80.6  | -80.6  |
| 5tys | -108.1 | -60.5  | -43.6  | -43.4  | -43.4  | -43.4  | -43.4  | -43.4  | -43.4  |
| 6d0d | -110.3 | -97.9  | -97.9  | -79.8  | -40.8  | -74.1  | 153.8  | -33.9  | 153.8  |
| 6d0e | -113.8 | -83.3  | -79.8  | -74.7  | -62.9  | -62.9  | -54.1  | -54.1  | -54.1  |
| 6uwb | -127.5 | -107.5 | -105.1 | -41.3  | -41.3  | -41.3  | -41.3  | -41.3  | -41.3  |
| 6uwc | -106.3 | -81.4  | -81.4  | -75.3  | -66.8  | -66.8  | -66.8  | -66.8  | -66.8  |
| 7upj | -74.0  | -70.7  | -60.9  | -50.1  | -37.5  | -37.5  | -37.5  | -37.5  | -37.5  |